\documentclass[twocolumn,aps,prl,showpacs,superscriptaddress,longbibliography,10pt]{revtex4-2} %
\usepackage[colorlinks=true,linkcolor=blue,citecolor=blue]{hyperref}
\usepackage{amsfonts}
\usepackage{subfigure}
\usepackage{amsmath}
\usepackage{mathrsfs}
\usepackage{amssymb}
\usepackage{amsbsy}
\usepackage{epsfig}
\usepackage{graphicx}
\usepackage{epstopdf}
\usepackage{mathdots}
\usepackage{color}
\usepackage{cleveref}
\usepackage{float}
\usepackage{graphicx}
\usepackage{rotating}
\usepackage{physics}
\usepackage{nicefrac}

\begin{document}
\title{Floquet engineering of point-gapped topological superconductors}

\author{Xiang Ji}
\affiliation{Department of Physics, Jiangsu University, Zhenjiang, 212013, China}

\author{Hao Geng}
\affiliation{Department of Physics, Jiangsu University, Zhenjiang, 212013, China}

\author{Naeem Akhtar}
\affiliation{Department of Physics, Jiangsu University, Zhenjiang, 212013, China}

\author{Xiaosen Yang} \altaffiliation{yangxs@ujs.edu.cn}
\affiliation{Department of Physics, Jiangsu University, Zhenjiang, 212013, China}

\date{\today}
\begin{abstract}
Non-Hermitian systems exhibit two distinct topological classifications based on their gap structure: line-gap and point-gap topologies. 
Although point-gap topology is intrinsic to non-Hermitian systems, its systematic construction remains a challenge. Here, we present the Floquet engineering approach for realizing point-gapped topological superconductors. By combining Floquet theory with particle-hole symmetry (PHS), we show that a point gap hosting robust Majorana edge modes emerges at the overlap of Floquet bands with opposite winding numbers. In the thermodynamic limit, even weak non-Hermiticity opens a point gap from a gapless spectrum, driving a topological phase transition and breaking non-Bloch parity-time ($\mathcal{PT}$) symmetry. This transition is accompanied by the appearance of the Floquet $Z_2$ skin effect. 
Furthermore, the point-gapped topological phase and the non-Bloch $\mathcal{PT}$ symmetry exhibit size-dependent phenomena driven by the critical skin effect. Our work offers a new pathway for exploring the point-gapped topological phases in non-Hermitian systems.

\end{abstract}
\maketitle

\section{I. Introduction}
Recently, non-Hermitian systems have garnered substantial interest for their unique topological phases~\cite{bender_real_1998,berry_physics_2004,rudner_topological_2009,hu_absence_2011,lee_anomalous_2016,hodaei_enhanced_2017,shen_topological_2018,yao_edge_2018,yao_nonhermitian_2018,kunst_biorthogonal_2018,koutserimpas_nonreciprocal_2018,lee_hybrid_2019,longhi_topological_2019,liu_secondorder_2019,zhu_secondorder_2019,luo_higherorder_2019,matsumoto_continuous_2020,xiao_nonhermitian_2020,ashida_nonhermitian_2020,bergholtz_exceptional_2021,ding_nonhermitian_2022,li_gainlossinduced_2022,zhang_review_2022,weidemann_topological_2022,okuma_nonhermitian_2023,lin_topological_2023,wang_nonhermitian_2024,jing_biorthogonal_2024}, which reveal a variety of phenomena absent in Hermitian systems.
A defining feature of non-Hermitian systems is their complex energy spectra, which support two distinct gap structures: the line gap, inherited from Hermitian systems~\cite{shen_topological_2018,yao_edge_2018,yao_nonhermitian_2018,kunst_biorthogonal_2018,koutserimpas_nonreciprocal_2018}, and the point gap, a unique non-Hermitian feature~\cite{gong_topological_2018, kawabata_symmetry_2019,lee_topological_2019,denner_exceptional_2021,nakamura_universal_2023,schindler_hermitian_2023,fang_pointgap_2023,nakamura_bulkboundary_2024,wan_quantumsqueezinginduced_2023}. 
Previous investigations have shown that the energy spectra of non-Hermitian systems differ significantly between periodic (PBC) and open-boundary conditions (OBC), leading to distinct gap structures~\cite{lee_anomalous_2016,yao_edge_2018,wan_quantumsqueezinginduced_2023}.
A nonzero winding number of a PBC energy spectrum correlates with the non-Hermitian skin effect under OBC~\cite{zhang_correspondence_2020,okuma_topological_2020,zhang_universal_2022}, where the bulk eigenstates are localized at the boundries of the system~\cite{yao_edge_2018,longhi_probing_2019,song_nonhermitian_2019,li_critical_2020}.
This phenomenon results in energy collapse and the breakdown of the traditional bulk-boundary correspondence~\cite{yao_nonhermitian_2018,kawabata_nonbloch_2020,yokomizo_nonbloch_2019,yang_nonhermitian_2020,wang_amoeba_2024}.

Based on their gap structures, non-Hermitian systems under OBC exhibit two distinct classes of topological phases: the line-gapped and point-gapped topological phases~\cite{gong_topological_2018, denner_exceptional_2021}.
Point-gapped topological phases, characterized by point-gap OBC spectrum and robust in-gap topological modes, are exclusive to non-Hermitian systems and introduce new avenues for topological classification~\cite{gong_topological_2018, kawabata_symmetry_2019}.
Despite the fundamental role of point-gap topology in non-Hermitian systems~\cite{liu_symmetry_2022,dash_floquet_2024}, a universal approach to realizing point-gapped topological states under OBC remains elusive.

In addition, periodic driving has emerged as a powerful tool for engineering highly tunable topological phases~\cite{oka_photovoltaic_2009,jiang_majorana_2011,lindner_floquet_2011,rudner_anomalous_2013,rechtsman_photonic_2013,wang_observation_2013,jotzu_experimental_2014,mahmood_selective_2016,eckardt_colloquium_2017}. 
The overlap of Floquet bands in such systems enables unique topological phenomena, such as topological $\pi$ modes, which have no static analogs~\cite{lindner_floquet_2011,rudner_anomalous_2013,cheng_observation_2019}. 
Periodically driven systems readily implement features such as gain/loss~\cite{li_lossinduced_2023} and non-reciprocal hopping~\cite{ke_floquet_2023}, providing a robust platform for exploring exotic non-Hermitian phenomena~\cite{gong_stabilizing_2015,huang_realizing_2016,zhan_detecting_2017,zhan_detecting_2017,chitsazi_experimental_2017,hockendorf_nonhermitian_2019,zhang_nonhermitian_2020,cao_nonhermitian_2021,ding_experimental_2021,zhou_dual_2021,zhou_nonhermitian_2023a,sun_photonic_2024,ji_generalized_2024}.

This paper provides a universal framework for realizing point-gapped topological phases in non-Hermitian systems by combining Floquet theory with particle-hole symmetry (PHS), as illustrated in Fig.~\ref{fig:fig1}(a). Due to the PHS, the PBC winding numbers of the particle and hole bands are opposite.
Consequently, when two Floquet bands overlap, they create a region with zero winding number, where OBC quasienergies do not collapse. This region in the complex quasienergy plane forms a point gap, providing a platform for point-gapped topological phases. We confirm the framework in a periodically driven non-Hermitian Kitaev chain. The non-Hermiticity induces a phase transition from a gapless, topologically trivial phase to a point-gapped, topologically nontrivial phase hosting robust Majorana edge modes. Moreover, the breaking of non-Bloch parity-time ($\mathcal{PT}$) symmetry~\cite{longhi_nonbloch_2019,xiao_observation_2021,weidemann_topological_2022} occurs simultaneously with the topological phase transition. Both the symmetry breaking and the point-gapped topological phase exhibit strong dependence on system size, driven by the critical skin effect. Finally, we verify the validity of our proposal in other typical non-Hermitian systems.

\section{II. Non-Hermiticity induces point-gap Majorana modes}
In particular, we introduce a periodically driven non-Hermitian Kitaev chain, whose Bloch Hamiltonian is given by
\begin{align}
H(k,t)=H_0(k) + \lambda \cos \omega t \sigma_z.
\label{eq:model}
\end{align}
The first term $H_0(k) = h_0(k) + \vec{h}(k) \cdot \vec{\sigma}$ denotes the Hamiltonian of a non-Hermitian Kitaev chain~\cite{kitaev_unpaired_2001} with $\vec{h}(k) = \big(0, h_y(k), h_z(k)\big)$, $h_0(k) = 2i\delta\sin k$, $h_y(k) = -2\Delta \sin k$, $h_z(k) = \mu + 2t\cos k$, and $\vec{\sigma}=(\sigma_{x},\sigma_{y},\sigma_{z})$ is the vector of Pauli matrices.
Here $\mu$, $\Delta$ and $t\pm\delta$ represent the strengths of the chemical potential, pairing, and left/right hopping, respectively. 
The non-Hermiticity is introduced by the non-reciprocal hopping with $\delta\neq 0$. 
$\lambda$ and $\omega$ represent the periodical driving strength and angular frequency, respectively.
The Hamiltonian is periodic in time, $H(k,t)=H(k,t+T)$ with $T=2 \pi/\omega$, and preserves the PHS: $\mathcal{C}H^{T}(k,t)\mathcal{C}^{-1}=-H(-k,t)$ with $\mathcal{C}=\sigma_{x}$.
In the frequency domain framework~\cite{lindner_floquet_2011,rudner_anomalous_2013}, the Schr\"odinger equation can be written as 
\begin{align}
\sum_{m^{\prime}} \mathcal{H}_{m, m^{\prime}}(k,\omega)\ket{\psi_{n, R}^{m^{\prime}}(k)}=\varepsilon_n(k)\ket{\psi_{n, R}^{m}(k)},
\end{align}
where $\varepsilon_n$ denotes the quasienergy of the $n$ Floquet band. $\mathcal{H}_{m, m^{\prime}}(k,\omega)= m\omega\delta_{m,m^{\prime}}+H_{m-m^{\prime}}(k,\omega)$ denotes the Floquet Hamiltonian with $H_m(k,\omega)=\frac{1}{T}\int_0^{T} \dd t H(k,t)e^{-im\omega t}$. 
The Floquet Hamiltonian satisfies $\mathcal{H}(k,\omega)=\mathcal{H}(k,n\omega)$ ($n\in\mathbb{Z}$) and preserves the PHS $C\mathcal{H}^{T}(k, \omega)C^{-1}=-\mathcal{H}(-k, \omega)$. 

\begin{figure}
\includegraphics[width=8.5cm]{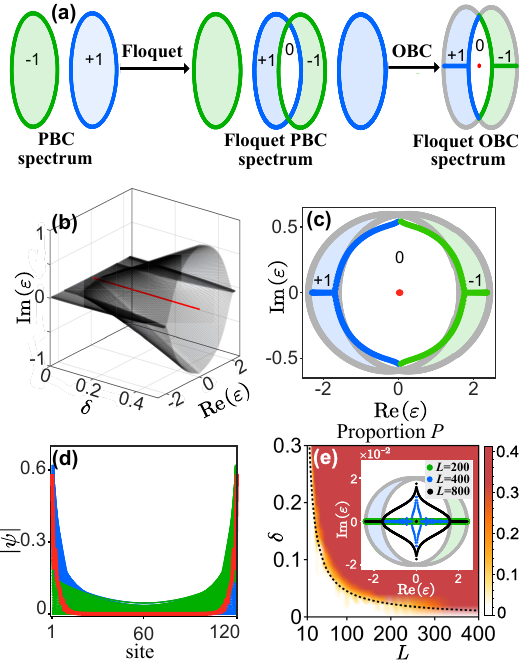}
\caption{(a) Schematic illustration of the mechanism of realizing the point-gapped topological phase in a periodically driven non-Hermitian systems. The regions with $+1$ and $-1$ PBC winding number are shadowed by blue and green, respectively.
(b) Quasienergy spectrum as a function of the non-Hermitian strength $\delta$ for an open periodically driven non-Hermitian Kitaev chain with $\mu=9.6$. The Floquet Majorana zero modes are highlighted in red.
(c) Complex quasienergy spectrum for $\delta=0.3$. The green (blue), and gray curves represent the OBC spectrum of particles (holes) and the PBC spectrum, respectively.
(d) Density distribution of eigenstates exhibit Floquet $Z_{2}$ skin effect.
(e) Complex quasieigenenergy proportion $P$ as a function of $\delta$ and lattice size $L$. Numerically, the quasienergies with $\Im(E)>10^{-8}$ are regarded as complex. 
The inset shows the complex quasienergy spectrum of the system with $\delta=0.01$ for different lattice sizes. 
Other parameters are $t=1$, $\Delta=0.8$, $\lambda=0.8$, and $\omega=10$.
\label{fig:fig1} } 
\end{figure}  

The quasienergy spectra of the Floquet Hamiltonian under OBC is depicted in Fig.~\ref{fig:fig1}(b).  In the Hermitian limit ($\delta=0$), the phase is topologically trivial with gapless OBC quasienergies. 
However, it is noticed that the OBC quasienergies are opened up with a point gap by the non-Hermiticity ($\delta\neq 0$). 
Notably, the point-gap phase hosts robust Majorana zero modes, which are encircled by the bulk quasienergies.
Figure~\ref{fig:fig1}(c) depicts the quasienergy spectra of the point-gapped topological superconductor with $\delta=0.3$ under PBC (gray curves) and OBC (green and blue dots) in the complex plane. 
Both the PBC and OBC quasienergies encircle the origin point and cannot be gapped by any reference line in the complex plane.
The Floquet Majorana zero modes, highlighted in red in Fig.~\ref{fig:fig1}(c), are surrounded by the OBC bulk quasienergies. 
This confirms that the phase is a Floquet point-gapped topological superconductor with robust Floquet Majorana zero modes, and these point-gapped topological phases are entirely induced by the non-Hermiticity. 
In addition, the essential difference between the two quasienergy spectra indeed indicates the emergence of the non-Hermitian skin effect, necessitating the non-Bloch band theory to fully describe the system~\cite{yao_edge_2018}.

The density distributions of the point-gapped topological superconductor are shown in Fig.~\ref{fig:fig1}(d). The eigenstates of the corresponding particle and hole are localized at the opposite end of an open chain, which exhibits a Floquet $Z_2$ skin effect.
The origin of the Floquet $Z_2$ skin effect can be revealed both by the winding number of the PBC spectra~\cite{okuma_topological_2020,zhang_correspondence_2020} and the non-Bloch band theory~\cite{yao_edge_2018,kawabata_nonbloch_2020,wang_amoeba_2024}. 
The winding number of the PBC spectra~\cite{okuma_topological_2020} can be defined on the Floquet Hamiltonian $\mathcal{H}(k)$, which is denoted as 
\begin{align}
W(\varepsilon)=\oint_0^{2\pi}\frac{\dd k}{2\pi i} \frac{\partial}{\partial_k} \log \det[\mathcal{H}(k,\omega)-\varepsilon].
\end{align}
The nonzero winding numbers of PBC quasienergies indicate the emergence of the Floquet non-Hermitian skin effect for a periodically driven system under OBC~\cite{okuma_topological_2020,zhang_correspondence_2020}. 
The winding numbers of the Floquet bands of particles and holes are $-1$ and $+1$, respectively. The opposite relationship is protected by the PHS:
\begin{align}
	W(\varepsilon)
	=&\oint_{0}^{2\pi}\frac{\dd k}{2\pi i}\frac{\partial}{\partial k}\log \det [\mathcal{H}(k,\omega)-\varepsilon]\nonumber\\
	=&\oint_{0}^{2\pi}\frac{\dd k}{2\pi i}\frac{\partial}{\partial k}\log \det [-C \mathcal{H}(-k,\omega)C^{-1}-\varepsilon]\nonumber\\
	=&-\oint_{0}^{2\pi}\frac{\dd k}{2\pi i}\frac{\partial}{\partial k}\log \det [ \mathcal{H}(-k,\omega)+\varepsilon]\nonumber\\
	=&-W(-\varepsilon).
\end{align}
For periodically driven systems, the winding number of different Floquet bands also satisfies $W(n\omega+\varepsilon)=W(\varepsilon)$, $n\in\mathbb{Z}$. Hence, we have
\begin{align}
W(n\omega/2+\varepsilon)=-W(n\omega/2-\varepsilon),\  n\in \mathbb{Z},
\label{eq:z2winding}
\end{align}
which indicates the Floquet $Z_2$ skin effect.

In systems exhibiting the non-Hermitian skin effect, the non-Bloch $\mathcal{PT}$ symmetry may depend on the system size~\cite{li_critical_2020}. To explore this, we introduce the complex quasienergy proportion, $P=N_c/N$, where $N_c$ is the number of complex quasienergies and $N$ is the total number of quasienergies. The variation of $P$ with respect to the non-Hermiticity strength $\delta$ and lattice size $L$ is shown in Fig.~\ref{fig:fig1}(e). For a fixed $\delta$, the phase has purely real quasienergy spectra, with $P=0$, when the system size is below a critical threshold $L_{c} \sim 1/\delta$. As the lattice size increases, the quasienergy spectra become complex, signaling the breaking of non-Bloch $\mathcal{PT}$ symmetry. This transition occurs at $L=L_c \sim 1/\delta$, meaning that even a weak non-Hermiticity can break the symmetry in the thermodynamic limit. Notably, the non-Bloch $\mathcal{PT}$ phase transition coincides with the topological phase transition. We confirm this double phase transition by examining the complex spectra of a long open chain with a small non-Hermiticity strength ($\delta=0.01$), as shown in the inset of Fig.~\ref{fig:fig1}(e).

\begin{figure}
\includegraphics[width=8.4cm]{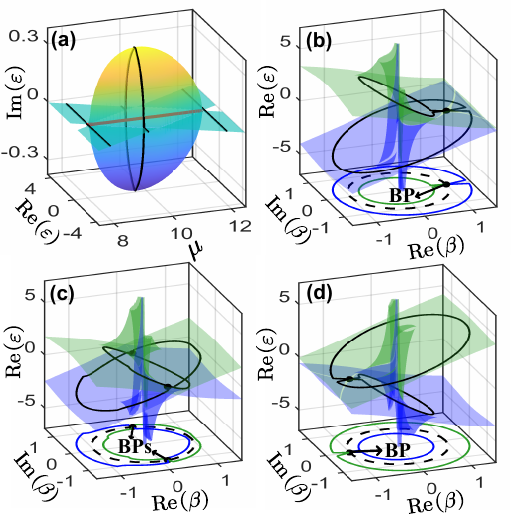}
\caption{(a) The OBC energy spectra as functions of $\mu$ with $\delta=0.3$. 
(b)-(d) The GBZ bands on the surface $f(\beta,\varepsilon)=0$ for $\mu = 8.0$, $9.6$ and $12.0$, respectively. The loops of particles ($C^p_{\beta}$) and holes ($C^h_{\beta}$) are split by the Brillouin zone (dotted circle), corresponding to $Z_{2}$ skin effect. The loops of particles and holes intersect at the BPs.}  \label{fig:fig2}
\end{figure}

\section{III. Double phase transitions}
To unveil the exotic double phase transition and the point-gapped Floquet topological superconductor, we present the three-dimensional OBC quasienergy spectra in Fig.~\ref{fig:fig2}(a). In the topologically nontrivial region, where $\mu \in  (8,12)$, Floquet Majorana zero modes emerge, surrounded by bulk quasienergies. Moreover, the topological phase transition aligns with the non-Bloch $\mathcal{PT}$ phase transition. Due to the non-Hermitian skin effect, we utilize non-Bloch band theory to fundamentally describe the novel phase and the double phase transition. The non-Bloch Floquet Hamiltonian $\mathcal{H}(\beta,\omega)$ is derived by generalizing the Bloch factor $e^{ik}$ to the non-Bloch factor $\beta=re^{ik}$, with $r$, $k\in \mathbb{R}$. The Hamiltonian satisfies PHS, where $C\mathcal{H}(\beta,\omega)C^{-1}=-\mathcal{H}(\beta^{-1},\omega)$. The values $\beta$ are determined by the characteristic equation:
\begin{align}
f(\beta,\varepsilon)=\det[\mathcal{H}(\beta,\omega)-\varepsilon]=0.
\label{eq:chq}
\end{align}
The number of the solution is even and can be ordered as $|\beta_{1}|\leqslant |\beta_{2}|\leqslant ... \leqslant |\beta_{2M}|$, $M\in \mathbb Z$.
The loops of the GBZ are formed by the trajectory of $\beta_M $ and $\beta_{M+1}$ with the continuum condition $\abs{\beta_M}=\abs{\beta_{M+1}}$~\cite{yao_edge_2018,yokomizo_nonbloch_2019,yang_nonhermitian_2020}.  Figure~\ref{fig:fig2}(b)-(d) show the GBZs for $\mu=8$, $\mu=9.6$, and $\mu=12$, respectively. The paired loops for particles ($C_{\beta}^{p}$) and holes ($C_{\beta}^{h}$) of the non-Bloch Floquet bands are denoted by green and blue curves and satisfy a reciprocal relationship:
\begin{align}
|\beta(n\omega/2+\varepsilon)| = |\beta(n\omega/2-\varepsilon)|^{-1},~ n\in \mathbb{Z},
\label{eq:fgbzPHS}
\end{align} 
which is protected by PHS and guarantees the emergence of the Floquet $Z_2$ skin effect in essence.
The gap closes when the reciprocal loops of the GBZs intersect at the BPs with $\beta_{c}=\pm1$ for an additional symmetry $\mathcal{H}^{*}(\beta,\omega) = \mathcal{H}(\beta^{*},\omega)$. This implies that the topological phase transition occurs at $\mu=8~ (12)$ when the gap closes at $\beta_{c}=1~ (-1)$, as shown in Fig.~\ref{fig:fig2}(b) and Fig.~\ref{fig:fig2}(d). At the topological phase transition, a single Bloch point(BP) splits into two conjugate BPs, marking the emergence of the point-gapped topological phase. Due to the PHS symmetry, the imaginary quasienergies are paired as $(\varepsilon_r, -\varepsilon_r)$, resulting in two conjugate BPs with $\beta(\varepsilon_r) =\beta^{*}(-\varepsilon_r)$ for the point-gapped topological superconductor, as shown in Fig.~\ref{fig:fig2}(c). Therefore, the appearance of the two conjugate BPs indicates both the topological phase transition and the breaking of non-Bloch $\mathcal{PT}$ symmetry. 

The topological phase transition can be analytically determined using the non-Bloch effective Hamiltonian derived from the non-Bloch Floquet operator $U(\beta, T)$:  
\begin{align}
U(\beta, T) = \mathbb{T} \left[ e^{-i\int^{T}_{0}{H}(\beta, t)dt} \right] \simeq e^{-iH_{eff}(\beta)T},
\end{align}
where $\mathbb{T}$ is the time-ordering operator. For the two-photon resonance case, the effective non-Bloch Hamiltonian takes the formula:
$H_{eff}(\beta) =  h_{0}(\beta) + \left[1-\frac{\omega}{h(\beta)} \right]\left[\vec{h}(\beta)+\frac{4h(\beta)}{3\omega}\cdot\vec{D}_{\bot} (\beta) \right]\cdot \vec{\sigma}$ with $\vec{D}_{\bot}(\beta)= \frac{\lambda h_{y}(\beta)}{h^{2}(\beta)}  \left(0, -h_{z}(\beta),  h_{y}(\beta)\right)$. The gap of the effective non-Bloch Hamiltonian closes when $\omega = h(\beta)$. By combining his gap closing condition at the BPs, we derive the topological phase boundaries as $\omega = |\mu \pm 2t|$, which agrees with the numerical results. 

\section{IV. Critical skin effect}
We have shown that both the point-gapped topological phase and the non-Bloch $\mathcal{PT}$ symmetry exhibit a size-dependent effect. 
\begin{figure}
	\includegraphics[width=8.5cm]{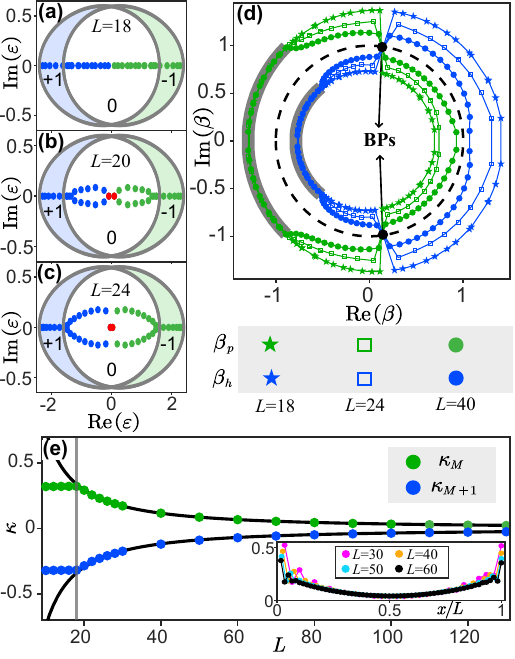}
	\caption{(a)-(c) Quasienergy spectra with $\mu=9.6$ for different lattice size. The green (blue) dots denote the OBC quasienergies of particles (holes), while the gray curves represent the PBC quasienergies. (d) The finite-size GBZ of particles ($\beta_p$) and holes ($\beta_h$) for systems with different lattice sizes. (e) The decay length $\kappa_i=-\log(\abs{\beta_i})$ as a function of lattice size for the points in finite-size GBZs, which are closest to the BP. The inset shows the corresponding density distribution.} \label{fig:fig3} 
\end{figure}
To explore this, we present the quasienergy spectra for various lattice sizes in Figs.~\ref{fig:fig3}(a)-(c). 
As the lattice size increases, a double phase transition occurs at a critical lattice size, $L_c=20$.
This size-dependent phenomenon arises from critical skin effect and can be revealed by finite-size GBZs~\cite{li_critical_2020,wang_scalefree_2023,li_universal_2024}, as shown in Fig.~\ref{fig:fig3}(d). 
One key feature of the GBZs is that the two conjugate BPs remain invariant to lattice size, regardless of the non-Bloch $\mathcal{PT}$ symmetry or topology. 
Another important observation is that, although the BPs are invariant, the GBZ curves near the BPs - corresponding to OBC quasienergies with zero PBC winding number, show a strong lattice-size dependence.  
As the size increases, these GBZs for particles and holes progressively approach each other and eventually meet at the BPs in the thermodynamic limit.
The scaling behavior of the decay lengths (or inverse localization lengths) $\kappa_i=-\log(\abs{\beta_i})$ for the $\beta$ nearest to the lower BP is shown in Fig.~\ref{fig:fig3}(e). The decay length scales with the lattice size as $\kappa \propto \pm c/L^{4/3}$, which is different from previous works where the scaling factor was $-1$~\cite{li_critical_2020,yokomizo_scaling_2021,wang_scalefree_2023}.
The eigenstate profile shows a scale-free exponential decay, as illustrated in the insert.
Additionally, the finite-size breaking of non-Bloch $\mathcal{PT}$ symmetry is linked to the formation of cusps in the finite-size GBZs~\cite{hu_geometric_2024}. 
As shown in Fig.~\ref{fig:fig3}(d), the shaded curves of the GBZ, corresponding to OBC quasienergies with non-zero PBC winding number, are independent of lattice size, with $\kappa=\pm 0.32$. As the lattice size increases, two paired cusps emerge in the finite-size GBZ, signifying branch points in the quasienergy spectrum and signaling the breaking of non-Bloch $\mathcal{PT}$ symmetry.

\section{IV. Fidelity susceptibility}
The biorthogonal fidelity susceptibility acts as an indicator for detecting exceptional points \cite{tzeng_hunting_2021}, signaling the occurrence of the double phase transition.
\begin{figure}
\includegraphics[width=8.1cm]{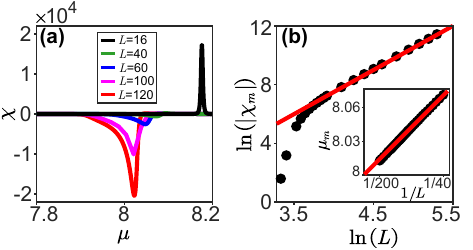}
\caption{(a) Ground state fidelity susceptibility $\chi$ varies with $\mu$ with different lattice size. (b) Finit-size scaling of the minimum of the susceptibility $\chi_{m} \propto - L^{3}$ . The insert shows the finit-size scaling of critical chemical potential.}  \label{fig:fig4}
\end{figure}
Here, we generalize the biorthogonal fidelity susceptibility~\cite{matsumoto_continuous_2020,sun_biorthogonal_2022,jing_biorthogonal_2024,liu_quantum_2024,ren_identifying_2024} to periodically driven systems, providing a useful method to identify such phase transition. The biorthogonal fidelity is defined in terms of the biorthogonal eigenstates of the Floquet Hamiltonian:
$F_n(\mu)= \sqrt{\bra{\psi^L_n(\mu)}\ket{\psi^R_n(\mu+\delta_\mu)}\bra{\psi^L_n(\mu+\delta_\mu)}\ket{\psi^R_n(\mu)} }$.
The fidelity susceptibility, which corresponds to the coefficient of the second derivative term in the Taylor expansion of the fidelity~\cite{sun_biorthogonal_2022}, can be written as
\begin{align}
\chi(\mu)=\lim_{\delta\mu\rightarrow 0}\frac{- 2 \ln\abs{F(\mu,\delta\mu)}}{(\delta\mu)^2}.
\end{align}
The ground state fidelity susceptibility of an open chain is depicted in Fig.~\ref{fig:fig4}(a). The susceptibility has a valley when the lattice size is larger than the critical size $L_c$.  Notably, the susceptibility tends towards $-\infty$ rather than $+\infty$ at the non-Bloch $\mathcal{PT}$ phase transition, as shown in Fig.~\ref{fig:fig4}(b). This negative susceptibility is caused by the metric difference in the Hilbert space~\cite{tzeng_hunting_2021}. The minimum of the susceptibility, $\chi_{m}$, follows a power law finite-size scaling: $\chi_{m} \propto - L^{3}$ near the phase transition. The finite-size scaling of $\mu_m$ with respect to $L$ is shown in the insert of Fig.~\ref{fig:fig4}(b). In the thermodynamic limit, $\mu_m$ approaches the double phase transition point. Thus, the double phase transition can indeed be detected by the fidelity susceptibility in future experiments. Moreover, the transition can also be detected through the dynamic behavior of wave package in the bulk~\cite{longhi_probing_2019,longhi_topological_2019}.

\section{V.Periodically driven $S$-wave topological superconductor}

The Floquet engineering of the point-gapped topological superconductors is independent of the specific form of the pairing and the number of bands. To confirm generality of this approach, we investigate a periodically driven non-Hermitian spinfull superconductor with  $s$-wave pairing, whose Bloch Hamiltonian takes the form:
\begin{align}
H_s(k,t)=& (2t\cos k -\mu)\tau_z + 2\alpha\sin k \tau_z\sigma_x +2i\delta \sin k \nonumber\\
 &+ h\tau_z\sigma_z-\Delta \tau_y\sigma_y+\lambda\cos(\omega t)\tau_z.
\label{eqHamSwave}
\end{align}
Here $\alpha$, $h$, and $\Delta$ denote the strength of spin-orbit
coupling, Zeeman field, and $s$-wave pairing, respectively. 
$\tau_{x/y/z}$ and $\sigma_{x/y/z}$ are the Pauli matrices that act on the particle-hole and spin degrees of freedom.
The Hamiltonian preserves the PHS, satisfying $\tau_xH_s^{T}(k,t)\tau_x=-H_s(-k,t)$.

\begin{figure}
\includegraphics[width=8.5cm]{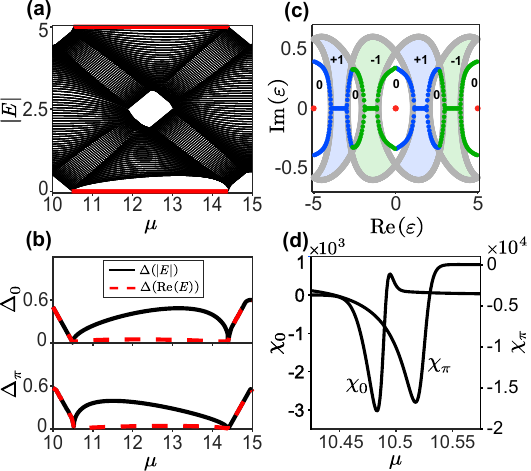}
\caption{(a) The OBC quasienergies as functions of $\mu$ with $\delta=0.3$. (b) The quasienergy gaps. (c) The OBC quasienergy spectrum for $\mu=11.5$.  (d) The fidelity susceptibilities of the states whose quasienergies are closest to 0 (denoted by $\chi_0$) and $\omega/2$ (denoted by $\chi_\pi$), respectively. Other parameters are $t=1, h=2.5$, $\alpha=0.8$, $\Delta=1.2$,  $\lambda=0.8$, $ \omega=10$. \label{fig5}}
\end{figure}

The OBC quasienergy spectrum as a function of $\mu$ is shown in Fig.~\ref{fig5}(a). 
Figure.~\ref{fig5}(b) shows the quasienergy gaps at $\varepsilon=0$ and $\varepsilon=\omega/2$. 
In the topological region, the system exhibits point gaps rather than real gaps in the quasienergy spectrum. The point gaps host Floquet  Majorana zero modes and $\pi$ modes, respectively, characterizing as a point-gapped topological phase.

To illustrate more clear, we show the quasienergy spectrum for $\mu=11.5$ in Fig.~\ref{fig5}(c), where point gaps are clearly visible at both $\varepsilon=0$ and $\varepsilon=\omega/2$. The Floquet engineering creates an overlap between different Floquet bands with opposite winding number, leading to a region with zero winding number and thereby giving rise to the point-gapped topological phase. 
The Floquet Majorana zero modes and $\pi$ modes are located within the gaps, encircled by the bulk spectrum. The topological phase transition also can be elucidated by the fidelity susceptibilities ($\chi_{0/\pi}$), shown in Fig.~\ref{fig5}(d), which detect the emergence of Floquet Majorana zero modes and $\pi$ modes.
In the thermodynamic limit, the fidelity susceptibility exhibits a clear valley at the topological phase transition, marking the onset of the point-gapped topological phase.

\section{V. Summary and discussion.}
In this paper, we have presented a general method for constructing point-gapped topological phases, paving the way for deeper exploration of topological non-Hermitian physics. Non-Hermiticity drives a phase transition from a gapless, topologically trivial phase to a point-gapped, topologically nontrivial phase with robust Floquet Majorana edge modes. This transition is accompanied by the breaking of non-Bloch $\mathcal{PT}$ symmetry, both of which exhibit strong dependence on system size due to the critical skin effect. 

In addition, higher-dimensional non-Hermitian systems exhibit a rich array of novel phenomena, such as higher-order topological states~\cite{liu_secondorder_2019,luo_higherorder_2019,zhu_secondorder_2019}, hybrid skin-topological effect~\cite{lee_hybrid_2019,li_gainlossinduced_2022}, and geometry-dependent skin effects~\cite{zhang_universal_2022,fang_geometrydependent_2022,wang_experimental_2023}. 
Point-gap topology plays a crucial role in understanding the topological properties of such systems and warrants further exploration. Our Floquet engineering approach can be naturally extended to higher-dimensional systems and provide a platform to study the novel properties of point-gapped phases.

\section{acknowledgments}
We thank Zhong Wang for fruitful discussions. This work is supported by Natural Science Foundation of Jiangsu Province (Grant No. BK20231320).

\bibliography{reference} 

\end{document}